\documentclass[11pt,a4paper,twoside]{article}      

\newcommand{\mechyear}{2007}
\newcommand{\mechvol}{37}

\usepackage[cp1251]{inputenc}
\usepackage[T2A]{fontenc}
\usepackage[russian]{babel}

\usepackage[dvips]{graphicx}
\usepackage[dvips]{color}

\usepackage{amsmath}
\usepackage{amssymb}
\usepackage{amsxtra}
\usepackage{latexsym}
\usepackage{ifthen}
\usepackage[centerlast,footnotesize,sl]{caption2}
\usepackage{bm}
\usepackage{exscale}

\usepackage{mtt37}

\usepackage[dvips]{graphicx}
\renewcommand{\mechvol}{39}
\renewcommand{\mechyear}{2009}

\newcommand{\R}{\mathbb R}
\newcommand{\rk}{\mathop {\rm rank}\nolimits}
\newcommand{\sgn}{\mathop {\rm sgn}\nolimits}
\newcommand{\rea}{\mathop {\rm Re}\nolimits}
\newcommand{\bd}{\boldsymbol}
\newcommand{\rmd}{d}
\newcommand{\rmi}{\mathrm{i}}
\newcommand{\bi}{\bf}

\begin{document}

\udk{531.38}

\title{БИФУРКАЦИОННЫЕ ДИАГРАММЫ \\ИНТЕГРАЛЬНЫХ ОТОБРАЖЕНИЙ ВОЛЧКА \\С СИНГУЛЯРНОЙ СИММЕТРИЕЙ}
      {Бифуркационные диаграммы с сингулярной симметрией}

\author{А.Ю.~Савушкин, И.И.~Харламова}

\date{10.11.09}

\address{Волгоградская академия гос. службы, Россия}

\email{sandro@vags.ru}

\maketitle

\begin{abstract}
В общем случае гамильтонова система с тремя степенями свободы,
описывающая движение твердого тела в поле двух постоянных сил, не
допускает групп симметрий. Х.\,Яхья нашел условия, при которых
уравнения движения гиростата Ковалевской в поле такого вида имеют, в
дополнение к интегралу энергии, интеграл, линейный по компонентам
угловой скорости. Позднее было отмечено, что в двойном силовом поле
этот интеграл при условиях Яхья существует для любого динамически
симметричного тела с центрами приложения полей в экваториальной
плоскости. Соответствующая система является натуральной механической
системой с $S^1$-симметрией, поэтому можно ставить вопрос о
реализации программы С.\,Смейла топологического анализа. В то же
время, эта симметрия обладает некоторым множеством особых точек и,
следовательно, не является регулярной. В настоящей работе строятся
бифуркационные диаграммы отображения момента для семейства систем с
сингулярной симметрией и исследуется зависимость диаграмм от
единственного существенного параметра -- отношения экваториального и
осевого моментов инерции.
\end{abstract}

\Section[n]{Введение}\label{sec1} Уравнения движения твердого тела
вокруг неподвижной точки в двух постоянных силовых полях (например,
гравитационном и магнитном) в подвижных осях имеют вид
\begin{equation}
\begin{array}{l}
\displaystyle{\frac{\rmd {\bi M}}{\rmd t}}
={\bi M} \times
{\boldsymbol \omega}  + {\bi r}_1 \times {\boldsymbol \alpha}+{\bi
r}_2 \times {\boldsymbol \beta},\\[3mm]
\displaystyle{\frac {\rmd {\boldsymbol \alpha}}{\rmd
t}}={\boldsymbol \alpha}\times {\boldsymbol \omega}, \qquad \frac
{\rmd {\boldsymbol \beta}}{\rmd t}={\boldsymbol \beta}\times
{\boldsymbol \omega}.
\end{array}
\label{eq1_1}
\end{equation}
Здесь ${\bi M}$ -- кинетический момент, ${\boldsymbol \omega}={\bi
M} {\bi I}^{-1}$ -- угловая скорость, ${\bi I}$ -- тензор инерции в
закрепленной точке $O$. Векторы ${\boldsymbol \alpha}, {\boldsymbol
\beta}$ представляют собой напряженности силовых полей. В
соответствии со второй группой уравнений (\ref{eq1_1}) эти векторы
неизменны в инерциальном пространстве. Постоянные в теле векторы
 ${\bi r}_1, {\bi r}_2$
есть радиус-векторы центров приложения полей. Для дальнейшего нам
удобно трактовать все векторы как строки, чем и обусловлена запись
тензора справа от вектора.

Ограничение системы (\ref{eq1_1}) на любой невырожденный уровень
$P^6(a,b,c)$ трех геометрических интегралов
\begin{equation}
|{\boldsymbol \alpha}|^2=a^2, \qquad|{\boldsymbol \beta}|^2=b^2,
\qquad {\boldsymbol \alpha} {\cdot} {\boldsymbol \beta}=c \qquad
(|c| < a b)\label{eq1_2}
\end{equation}
в пространстве ${\R}^9({\boldsymbol \alpha}, {\boldsymbol
\beta},{\bi M})$ является гамильтоновой системой с тремя степенями
свободы по отношению к скобке Ли~--~Пуассона \cite{Bogo}, для
которой геометрические интегралы служат функциями Казимира. Функция
Гамильтона такова
\begin{equation}
H=\frac{1}{2}{\bi M}\cdot {\boldsymbol \omega}-{\bi r}_1 \cdot
{\boldsymbol \alpha}-{\bi r}_2 \cdot {\boldsymbol
\beta}.\label{eq1_3}
\end{equation}

Пусть $O{\bi e}_1 {\bi e}_2 {\bi e}_3$ -- ортонормированный базис
главных осей инерции. Предположим, что главные моменты инерции
удовлетворяют отношению Ковалевской 2:2:1, а векторы ${\bi r}_1,
{\bi r}_2$ параллельны экваториальной плоскости $O{\bi e}_1 {\bi
e}_2$ и образуют ортонормированную пару. Тогда, очевидно, можно
полагать без ограничения общности, что ${\bi e}_1={\bi r}_1, {\bi
e}_2={\bi r}_2$. При этих условиях в работе \cite{Bogo} в дополнение
к $H$ найден первый интеграл $K$, обобщающий интеграл Ковалевской.
Если в то же время
\begin{equation}\label{eq1_4}
a=b, \qquad c=0,
\end{equation}
то имеет место случай Яхья \cite{Yeh}; система допускает
$S^1$-симметрию и поэтому приводима к семейству интегрируемых систем
с двумя степенями свободы. Ввиду особенностей действия группы
приведенные системы определены не глобально. Интеграл момента
(гамильтониан действия группы) имеет вид
\begin{equation}\label{eq1_5}
L= {\bi M}\cdot ({\boldsymbol \gamma} - a^2 {\bi e}_3),
\end{equation}
где ${\boldsymbol \gamma}={\boldsymbol \alpha}\times {\boldsymbol
\beta}$.

На самом деле, как показано в \cite{KhRCD1}, интеграл (\ref{eq1_5})
имеет более общую природу. Пусть тензор инерции имеет ось симметрии
$O{\bi e}_3$, отношение экваториального момента инерции к осевому
произвольно, радиус-векторы центров приложения также произвольны с
одним лишь условием параллельности экваториальной плоскости
\begin{equation}\label{eq1_6}
{\bi r}_1 \cdot {\bi e}_3=0, \qquad {\bi r}_2 \cdot {\bi e}_3=0.
\end{equation}
Пусть $D$ -- невырожденная $2\times2$-матрица. Преобразование
\begin{equation*}
\left\|
\begin{array}{c} {\bi r}_1 \\ {\bi r}_2 \end{array}\right\|
\mapsto D \left\|
\begin{array}{c} {\bi r}_1 \\ {\bi r}_2 \end{array}\right\|, \qquad
\left\|
\begin{array}{c} {\boldsymbol \alpha} \\ {\boldsymbol \beta} \end{array}\right\|
\mapsto {(D^{-1})^T} \left\|
\begin{array}{c} {\boldsymbol \alpha} \\ {\boldsymbol \beta}
\end{array}\right\|, \qquad {\bi M} \mapsto {\bi M}
\end{equation*}
сохраняет уравнения (\ref{eq1_1}) и функцию (\ref{eq1_3}), приводя к
эквивалентной системе. Постоянные симметричные матрицы
\begin{equation*}
R=\left\|
\begin{array}{cc} {\bi r}_1 \cdot {\bi r}_1 &  {\bi r}_1 \cdot {\bi r}_2 \\
{\bi r}_2 \cdot {\bi r}_1 &  {\bi r}_2 \cdot {\bi r}_2
\end{array}\right\|,\qquad A=\left\|\begin{array}{cc} {\boldsymbol
\alpha} \cdot {\boldsymbol \alpha} &
{\boldsymbol \alpha} \cdot {\boldsymbol \beta} \\
{\boldsymbol \beta} \cdot {\boldsymbol \alpha} & {\boldsymbol \beta}
\cdot {\boldsymbol \beta}
\end{array}\right\|^{-1}
\end{equation*}
преобразуются по закону $R \mapsto D R D^T$,  $A \mapsto D A D^T$.
Существует $D\in GL(2,\R)$ такая, что $R$ становится единичной, а
$A$ -- диагональной ($c=0$). Очевидно, свойство (\ref{eq1_6})
сохраняется и для новой, но уже ортонормированной пары $\bi r_1, \bi
r_2$. Следовательно, она может быть выбрана в качестве главного
базиса инерции в экваториальной плоскости
\begin{equation*}
{\bi e}_1 = {\bi r}_1, \qquad {\bi e}_2 = {\bi r}_2.
\end{equation*}
Таким образом, изначальное требование ортонормированности
радиус-векто\-ров центров приложения и ортогональности полей
является излишним, так как система всегда приводится линейной
заменой переменных с постоянными коэффициентами к системе, у которой
пара ${\bi r}_1, {\bi r}_2$ ортонормирована, а ${\bd \alpha}, {\bd
\beta}$ -- ортогональна. Такая замена впервые предложена в работе
\cite{Kh34} и известна как параметрическая редукция для двух
постоянных полей.

Предположим, что {\it после} редукции получено условие $a=b$. Тогда
выполнены условия (\ref{eq1_4}) без дополнительных ограничений на
моменты инерции. Пусть
\begin{equation*}
T(\tau)=\left\| \begin{array}{ccc} \cos\tau & \sin\tau & 0 \\
-\sin\tau & \cos\tau & 0 \\ 0 & 0 & 1
\end{array}\right\|.
\end{equation*}
Однопараметрическое действие $g_\tau: {\R}^9({\boldsymbol \alpha},
{\boldsymbol \beta},{\bi M}) \to {\R}^9({\boldsymbol \alpha},
{\boldsymbol \beta},{\bi M})$, определенное как
\begin{equation}\label{eq1_7}
g_\tau(\left\|\begin{array}{c}{\bd \alpha} \\ {\bd \beta}\\{\bi
M}\end{array} \right\|) = T(\tau)\left\|\begin{array}{c}{\bd \alpha}
\\ {\bd \beta}\\{\bi M}\end{array} \right\|T(-\tau),
\end{equation}
сохраняет систему (\ref{eq1_1}), гамильтониан $H$, а также имеет
инвариантное многообразие $P^6(a,a,0)$. Следовательно,
$\{g_\tau\}\cong S^1$ -- группа симметрий. Соответствующий
циклический интеграл совпадает с (\ref{eq1_5}) \cite{KhRCD1}.

Обозначим через $n$ отношение экваториального момента инерции к
осевому. Единицы изменения выберем так, чтобы ${\bi I}=\mathop{\rm
diag}\nolimits \{n,n,1\}$, $a=1$. Далее пространство $P^6(1,1,0)$
обозначаем для краткости через $P^6$. Первые интегралы уравнений
(\ref{eq1_1}) на $P^6$ таковы
\begin{equation*}
\begin{array}{l} \displaystyle{H=\frac{1}{2}\big[n
(\omega_1^2+\omega_2^2)+\omega_3^2\big]-\alpha_1 -\beta_2,}\\[3mm]
L=n[\omega_1 (\alpha_2 \beta_3-\alpha_3 \beta_2)+\omega_2 (\alpha_3
\beta_1-\alpha_1 \beta_3)]+\omega_3(\alpha_1 \beta_2-\alpha_2
\beta_1-1).
\end{array}
\end{equation*}

Обозначим матрицу со строками ${\bd \alpha}$, ${\bd \beta}$, ${{\bd
\alpha} \times {\bd \beta}}$ через~$Q$. Очевидно, отображение
$({\boldsymbol \alpha}, {\boldsymbol \beta},{\bi M}) \mapsto (Q,{\bd
\omega})$ есть диффеоморфизм $P^6$ на $T SO(3)$. Система
(\ref{eq1_1}), ограниченная на $P^6$, является поэтому натуральной
механической системой на $SO(3)$ с $S^1$-симметрией в смысле
\cite{Smale}. Программа Смейла топологического анализа этой системы
может быть выполнена с некоторыми модификациями, учитывающими тот
факт, что действие группы обладает множеством неподвижных точек
${\bd \alpha} \times {\bd \beta}={\bi e}_3$ (однопараметрическая
подгруппа действует на $SO(3)$ внутренними автоморфизмами, поэтому
она сама, будучи коммутативной, является множеством неподвижных
точек). В настоящей работе решается задача вычисления бифуркационной
диаграммы $\Sigma$ отображения момента
\begin{equation*}
J=L\times H: P^6 \to \R^2.
\end{equation*}
Также предъявляются различные типы диаграмм $\Sigma$ в зависимости
от параметра $n$. Напомним, что по определению $\Sigma$ состоит из
точек $(\ell,h) \in \R^2$, над окрестностями которых отображение $J$
не является локально тривиальным. Поэтому интегральные многообразия
\begin{equation*}
J_{\ell,h}=\{\zeta \in P^6: L(\zeta)=\ell, H(\zeta)=h\}
\end{equation*}
претерпевают топологические перестройки, когда $(\ell,h)$ пересекает
$\Sigma$. В частности, нахождение множества $\Sigma$ является
необходимым этапом топологического анализа задачи. Диаграмма
$\Sigma$ совпадает с множеством критических значений $J$ ввиду
компактности изоэнергетических уровней.

\vskip2mm\Section{Критические точки первых интегралов}\label{sec2}
Воспользуемся заменой переменных, введенной в \cite{Kh32} и
обобщающей замену С.\.Ковалевской на случай двух силовых полей
($\rmi^2=-1$):
\begin{equation}\label{eq2_1}
\begin{array}{ll}
x_1 = (\alpha_1  - \beta_2) + \rmi (\alpha_2  + \beta_1),&
x_2 = (\alpha_1  - \beta_2) - \rmi(\alpha_2  + \beta_1 ), \\
y_1 = (\alpha_1  + \beta_2) + \rmi(\alpha_2  - \beta_1), & y_2 =
(\alpha_1  + \beta_2) -
\rmi(\alpha_2  - \beta_1), \\
z_1 = \alpha_3  + \rmi\beta_3, &
z_2 = \alpha_3  - \rmi\beta_3,\\
w_1 = \omega_1  + \rmi\omega_2 , & w_2 = \omega_1  - \rmi\omega_2, \\
w_3 = \omega_3.&
\end{array}
\end{equation}
Условия (\ref{eq1_2}) на $P^6$ принимают вид
\begin{equation}\label{eq2_2}
z_1^2+x_1 y_2=0, \qquad z_2^2+x_2 y_1=0, \qquad x_1 x_2+y_1 y_2+2z_1
z_2=4.
\end{equation}
Введем переменные $x,y,z$, полагая
\begin{equation*}
x^2=x_1 x_2,\qquad y^2=y_1 y_2, \qquad z^2=z_1 z_2,
\end{equation*}
и примем следующее соглашение о знаках:
\begin{equation}\label{eq2_3}
x \geqslant 0, \qquad \sgn y = \sgn \rea (y_i).
\end{equation}
Тогда
\begin{equation}\label{eq2_4}
z^2=\pm x y, \qquad (x \pm y)^2=4, \qquad x \in [0,2].
\end{equation}

При исследовании критических точек различных функций на $P^6$ для
того, чтобы избежать введения неопределенных множителей Лагранжа для
ограничений (\ref{eq2_2}), удобно использовать уравнения,
предложенные в работе \cite{KhJPA}.
\begin{lemma}
Пусть $f$ -- гладкая функция комплексных переменных $(\ref{eq2_1})$.
Критические точки ограничения функции $f$ на подмногообразие,
определенное уравнениями $(\ref{eq2_2})$, описываются системой
уравнений
\begin{equation}\label{eq2_5}
\begin{array}{l}
\partial_{w_1} f=0, \qquad \partial_{w_2} f=0, \qquad \partial_{w_3} f=0, \\
(2 z_2 {\partial_{x_2} } + 2 z_1 {\partial_{y_2} } -x_1
{\partial_{z_1} }  - y_1 {\partial_{z_2} })f=0, \\
(2 z_1 {\partial_{x_1} } + 2 z_2 {\partial_{y_1} } -x_2
{\partial_{z_2} }  - y_2 {\partial_{z_1} })f=0, \\
(x_1 {\partial_{x_1}} - x_2 {\partial_{x_2} } + y_1 {\partial_{y_1}}
- y_2 {\partial_{y_2}})f=0.
\end{array}
\end{equation}
\end{lemma}

Пусть $\mathcal{C}$ -- множество критических точек интегрального
отображения $J$. Тогда ${\mathcal {C}= \mathcal {C}^0 \cup \mathcal
{C}^1}$, где ${\mathcal{C}^i=\{\zeta \in P^6: \rk J(\zeta)=i\}}$.

Вначале рассмотрим критические точки гамильтониана $H$. Они являются
положениями равновесия в системе (\ref{eq1_1}). Уравнения
(\ref{eq2_5}) с $f=H$ дают
\begin{equation*}
w_1=w_2=w_3=0,\qquad z_1=z_2=0, \qquad y_1=y_2.
\end{equation*}
Из (\ref{eq2_4}) следует, что в этом случае
\begin{equation}\label{eq2_6}
x y =0.
\end{equation}
Если $x=0$, то $y_1=y_2=\pm 2$. Получаем два положения равновесия
\begin{equation*}
\begin{array}{lll}
{\bd \omega=0}, & {\bd \alpha}={\bi e}_1, & {\bd \beta}={\bi
e}_2;\\
{\bd \omega=0}, & {\bd \alpha}=-{\bi e}_1, & {\bd \beta}=-{\bi e}_2.
\end{array}
\end{equation*}
Легко проверить, что оба они невырождены, первое -- устойчиво,
второе -- неустойчиво. Соответствующие значения первых интегралов
дают две точки в плоскости $(\ell,h)$: $P_-(0,-2)$ и $P_+(0,2)$.
Если в (\ref{eq2_6}) взять $y=0$, то значения $x_1,x_2$ остаются
произвольными в пределах условия $x_1 x_2=4$. Следовательно, имеется
целая окружность вырожденных безразличных положений равновесия,
отвечающая точке $P_0(0,0)$ в плоскости ${(\ell,h)}$. С физической
точки зрения это множество положений равновесия состоит из всех
ориентаций тела, в которых экваториальная плоскость совпадает с
плоскостью напряженностей сил $O{\bd \alpha}{\bd \beta}$ и последний
базис имеет противоположную ориентацию с базисом $O{\bi e}_1 {\bi
e}_2$. При этом вращающий момент сил ${\bi e}_1\times{\bd
\alpha}+{\bi e}_2\times{\bd \beta}$ тождественно равен нулю (все эти
утверждения о положениях равновесия можно вывести также из
результатов работы \cite{HasYeh}, где рассматривается случай трех
постоянных полей и его вырождения).

Найдем критические точки интеграла $L$, записав уравнения
(\ref{eq2_5}) с $f=L$:
\begin{equation*}
\begin{array}{l}
x_2 z_1-y_2 z_2=0,\qquad y_1 z_1-x_1 z_2=0,\qquad
x_1 x_2 -y_1 y_2=-4, \\
2 n w_1 + (y_1 z_1-x_1 z_2)w_3=0, \qquad  2 n w_2 + (y_2 z_2-x_2
z_1)w_3=0.
\end{array}
\end{equation*}
С учетом (\ref{eq2_2}) получим
$$
w_1=w_2=0, \qquad z_1=z_2=0, \qquad x_1 =x_2 =0, \qquad y_1 y_2=4.
$$
Пусть $y_1=2 \exp (-\rmi \psi)$, $y_2=2 \exp (\rmi \psi)$. Тогда
из (\ref{eq2_1}), (\ref{eq1_1}) находим
\begin{equation}\label{eq2_7}
\begin{array}{l}
{\bd \alpha}={\bi e}_1\cos \psi-{\bi e}_2\sin \psi,\quad
{\bd \beta}={\bi e}_1\sin \psi+{\bi e}_2\cos \psi,\\
{\omega}_1={\omega}_2=0,\quad {\omega}_3=\dot \psi, \qquad \ddot
\psi = -2 \sin \psi.
\end{array}
\end{equation}
Эти уравнения описывают множество точек, принадлежащих траекториям
маятниковых движений около оси $O{\bd \gamma}=O{\bi e}_3$, а
последнее уравнение определяет соответствующую квадратуру. Значения
первых интегралов таковы:
\begin{equation*}
\ell =0, \qquad \displaystyle{h=\frac{1}{2}\omega_3^2-2 \cos \psi
\geqslant -2.}
\end{equation*}
Отметим, что множество (\ref{eq2_7}) включает и два невырожденных
положения равновесия, но не содержит вырожденные равновесия ${\bd
\gamma} = -{\bi e}_3$. Поэтому в точках вырожденных равновесий $dL
\ne 0$, и, следовательно, множество $\mathcal{C}^0$ состоит в
точности из двух точек пространства $P^6$. Естественно ожидать, что
существуют нетривиальные движения с ${\bd \gamma} \equiv -{\bi
e}_3$. Такое множество, если оно существует, не является критическим
ни для одного из интегралов $H,L$. Эта возможность рассматривается в
следующем параграфе.

\vskip2mm\Section{Критические движения общего вида и значения
интегралов}\label{sec3} Точки множества $\mathcal{C}$, не
являющиеся критическими ни для одного из интегралов $H$ или $L$,
порождаются критическими движениями общего вида. Эти движения --
периодические решения системы (\ref{eq1_1}), являющиеся
одновременно орбитами действия $g_\tau$, у которых $\tau = \sigma
t$ ($\sigma$ -- некоторая константа). Следовательно, выражения
(\ref{eq1_7}) с такой зависимостью $\tau(t)$ дают аналитическое
решение для движений этого типа. Соответствующая часть множества
$\mathcal{C}$ описывается системой (\ref{eq2_5}) при $f=H-\sigma
L$. Первые три уравнения дают
\begin{equation}\label{eq3_1}
\begin{array}{c}
\displaystyle{w_1=-\frac{1}{2}(y_1 z_1 -x_1 z_2)\sigma,}\qquad
\displaystyle{w_2=-\frac{1}{2}(y_2 z_2 -x_2 z_1)\sigma,},\\
\displaystyle{w_3=-\frac{1}{4}(x^2-y^2+4)\sigma.}
\end{array}
\end{equation}

Для исходных переменных эти равенства выражают тот факт, что
движения служат орбитами группы симметрий, т.е. пропорциональность
угловой скорости движения угловой скорости вдоль орбиты действия
группы:
\begin{equation*}
{\bd \omega} = \sigma ({\bd \gamma} - {\bf e}_3).
\end{equation*}

Исключая $w_j$ с помощью (\ref{eq3_1}) из уравнений второй группы
полученной из (\ref{eq2_5}) системы, находим
\begin{eqnarray}
& y_1=y_2,\label{eq3_2}
\end{eqnarray}
\begin{equation}
\begin{array}{c}
 x_1 z_2 u-(y_1 u-8)z_1=0, \\[1mm]
 x_2 z_1 u-(y_2 u-8)z_2=0.\label{eq3_3}
\end{array}
\end{equation}
Для сокращения записи обозначено
\begin{equation}\label{eq3_4}
u=\big[4-(n-1)(x^2-y^2)\big]\sigma^2.
\end{equation}
Отметим, что в соответствии с (\ref{eq3_2}) и (\ref{eq2_3}) здесь
$y= y_1=y_2$.

Система (\ref{eq3_3}) линейна и однородна относительно $z_1,z_2$.
Предположим вначале, что $z_1=z_2=0$. Из (\ref{eq3_1}) получаем
$w_1=w_2=0$, а тогда из (\ref{eq2_4}) следует, что выполнено одно из
двух: либо $x=0$, либо $y=0$. Если $x=0$, то $y^2=4$, и последнее
уравнение (\ref{eq3_1}) дает $w_3=0$. Это два невырожденных
равновесия, изученных выше. В свою очередь, если положить $x\ne 0$,
то $y=0$, $x=2$. Компонента $w_3$ остается произвольной. Для
переменных в системе (\ref{eq1_1}) имеем ${\bd \gamma}=-{\bi e}_3$,
${\bd \omega}=2 \sigma {\bi e}_3$. Эти движения являются
равномерными вращениями вокруг третьей оси инерции, которая остается
ортогональной плоскости сил, в то время как базисы $O{\bi e}_1 {\bi
e}_2$ и $O{\bd \alpha}{\bd \beta}$ задают противоположную ориентацию
в этой плоскости. Значения первых интегралов $h=2\sigma^2$,
$\ell=4\sigma$ заполняют параболу $h=\ell^2/8$.

Рассмотрим теперь случай $z_1 z_2 \ne 0$. Выразим первые интегралы в
переменных~(\ref{eq2_1}):
\begin{equation*}
\begin{array}{l} \displaystyle{L=\frac{n}{4}\big[(x_2 z_1-y_2 z_2)
w_1+(x_1 z_2-y_1
z_1)w_2\big]-\frac{1}{4}(x^2-y^2+4)w_3,}\\[3mm]
\displaystyle{H=\frac{1}{2}(n w_1 w_2
+w_3^2)-\frac{1}{2}(y_1+y_2)}.\\
\end{array}
\end{equation*}
Тогда из (\ref{eq3_1}), (\ref{eq3_2}), (\ref{eq2_4}) находим
значения
\begin{equation}\label{eq3_5}
\begin{array}{l} \displaystyle{\ell =
\frac{\sigma}{16}\{16+8[(n+1)x^2+(n-1)y^2]-(2n-1)(x^2-y^2)^2\}},\\
\displaystyle{h = -y +\frac{\sigma}{2}\ell,}\\
y=\pm (2-x), \qquad x \in [0,2].
\end{array}
\end{equation}
Ненулевые решения системы (\ref{eq3_3}) по $z_1,z_2$ существуют,
если
\begin{equation*}
[(x+y)u-8][(x-y)u+8]=0.
\end{equation*}
Отсюда, подставляя значение $u$ из (\ref{eq3_4}), получаем
\begin{equation*}
\sigma^2=\frac{\sgn y}{n-(n-1)x}\quad \text{или}
\quad\sigma^2=\frac{\sgn y}{(1-x)[n-(n-1)x]}.
\end{equation*}
Эти выражения вместе с (\ref{eq3_5}) определяют значения $\ell, h$
на бифуркационной диаграмме как функции от одного параметра $x$.
Допустимые значения $x$ вырезаются из базового отрезка $[0,2]$
соответствующим условием $\sigma^2(x) \geqslant 0$.

\vskip2mm\Section{Бифуркационная диаграмма}\label{sec4} Обозначим
\begin{equation}\label{eq4_1}
\begin{array}{ll}
\varphi_0(x)=x[2n-(n-1)x],  \\[2mm]
\varphi_1(x)=n-(n-1)x, & \varphi_2(x)=(1-x)\varphi_1(x),  \\[2mm]
\displaystyle{h_1(x)=-\frac{5}{2}+\frac{3}{2}x+\frac{n+x}{2\varphi_1(x)},}&
\displaystyle{h_2(x)=-\frac{5}{2}+x+\frac{n+x}{2\varphi_2(x)}.}
\end{array}
\end{equation}
Следующая теорема суммирует сказанное выше.
\begin{theorem}
Бифуркационная диаграмма $\Sigma$ отображения момента для
динамически симметричного волчка в $S^1$-симметричной паре
постоянных силовых полей состоит из следующих подмножеств плоскости
$(\ell,h)$:
\begin{equation*}
\begin{array}{l}
\displaystyle{\delta_0=\{P_-,P_+,P_0\},}\qquad
\displaystyle{\delta_1=\{\ell=0 : h \geqslant -2\},}\\[2mm]
\displaystyle{\delta_2=\{h=\frac{1}{8}\ell^2 : \ell \in \R\},}\\
\displaystyle{\delta_3=\{\ell= \pm \frac{\varphi_0(x)}{\sqrt{\varphi_1(x)}}, h=h_1(x) : x\in I_3\},}\\
\displaystyle{\delta_4=\{\ell= \pm \frac{\varphi_0(x)}{\sqrt{\varphi_2(x)}}, h=h_2(x) : x\in I_4\},}\\
\displaystyle{\delta_5=\{\ell= \pm \frac{\varphi_0(x)}{\sqrt{-\varphi_1(x)}}, h=-h_1(x) : x\in I_5\},}\\
\displaystyle{\delta_6=\{\ell= \pm \frac{\varphi_0(x)}{\sqrt{-\varphi_2(x)}}, h=-h_2(x) : x\in I_6\},}\\
\end{array}
\end{equation*}
где
\begin{equation}\label{eq4_2}
\begin{array}{ll}
 I_3= \left\{ \begin{array}{ll}
[0,2], & n < 2 \\[2mm]
\left[0, \displaystyle{\frac{n}{n-1}}\right], & n \geqslant  2
\end{array}\right., \quad & I_4= \left\{ \begin{array}{ll}
[0,2), & n \leqslant 2 \\[2mm]
[0, 1) \cup \left(\displaystyle{\frac{n}{n-1}},2\right], & n > 2
\end{array}\right.,\\[7mm]
I_5= \left\{ \begin{array}{ll}
\emptyset, & n \leqslant 2 \\[2mm]
\left(\displaystyle{\frac{n}{n-1}},2\right], & n >  2
\end{array}\right., \quad & I_6= \left\{ \begin{array}{ll}
(1,2], & n < 2 \\[2mm]
\left(1,\displaystyle{\frac{n}{n-1}}\right), & n \geqslant 2
\end{array}\right..
\end{array}
\end{equation}
\end{theorem}

\begin{remark} Очевидно, что $\delta_0
\subset \delta_1$. Однако мы выделяем трехточечное множество
$\delta_0$ как порожденное состояниями равновесия тела. Отметим
также, что параметр $x$ на кривых \mbox{$\delta_3$ -- $\delta_6$}
равен значению $\sqrt{(\alpha_1-\beta_2)^2+(\alpha_2+\beta_1)^2}$,
которое, тем самым, оказывается постоянным вдоль любой критической
траектории.\end{remark}

\vskip2mm\Section{Примеры диаграмм и зависимость от физического
параметра}\label{sec5} Рассматриваемая система, ее отображение
момента и бифуркационная диаграмма зависят от безразмерной
характеристики $n$, выражающей отношение экваториального момента
инерции к осевому. Из выражений (\ref{eq4_2}) для сегментов
изменения $x$ вдоль бифуркационных кривых следует, что значение
$n=2$, отвечающее случаю Яхья, разделяет принципиально различные
типы диаграмм. Нетрудно проверить аналитически, что при $n<2$
диаграммы не претерпевают качественных изменений. Даже в случае
$n=1$ при наличии некоторых очевидных вырождений в выражениях
(\ref{eq4_1}), диаграмма топологически устроена так же, как и при
близких значениях $n$. Типичная диаграмма для $0<n<2$ показана на
рис.~1 (в силу симметрии относительно оси $h$ иллюстрируется
только часть $\ell \geqslant 0$).

\begin{figure}[ht]
\centering
\includegraphics[width=45mm,keepaspectratio]{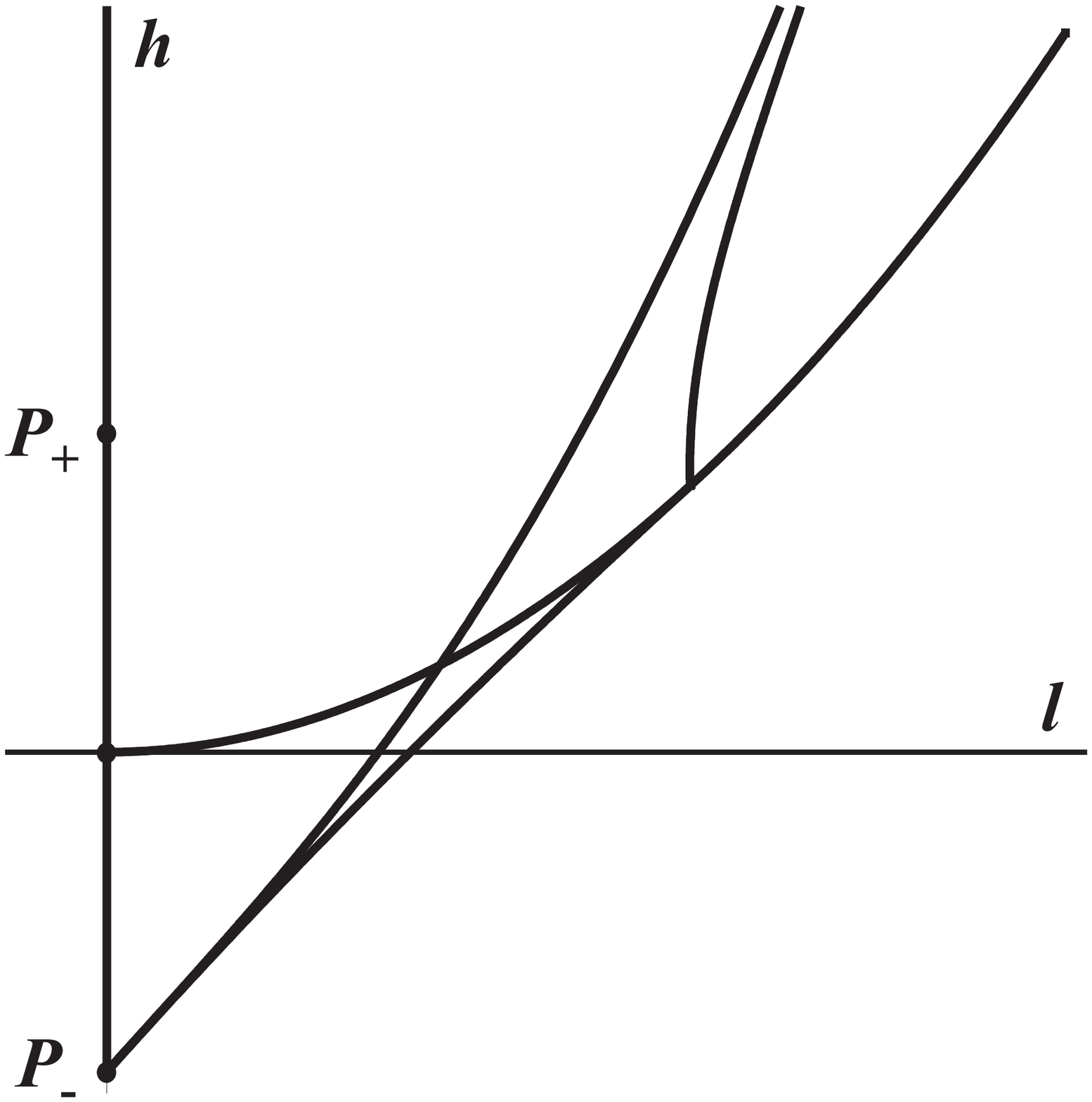}
\caption{Бифуркационная диаграмма при $n<2$.}
\end{figure}

Для значений $n>2$ существует несколько типов бифуркационных
диаграмм. Они различаются по количеству узлов (точек пересечения
гладких отрезков диаграммы, точек возврата и т.п.) и ячеек
регулярности (связных компонент множества $\R^2 \backslash
\Sigma$). Две диаграммы для случая $n>2$ вместе с увеличенными
фрагментами показаны на рис.~2. Здесь для примера взяты значения
\emph{а}) $n=2.3$; \ \emph{б}) $n=4$.

\begin{figure}[ht]
\centering
\includegraphics[width=60mm,keepaspectratio]{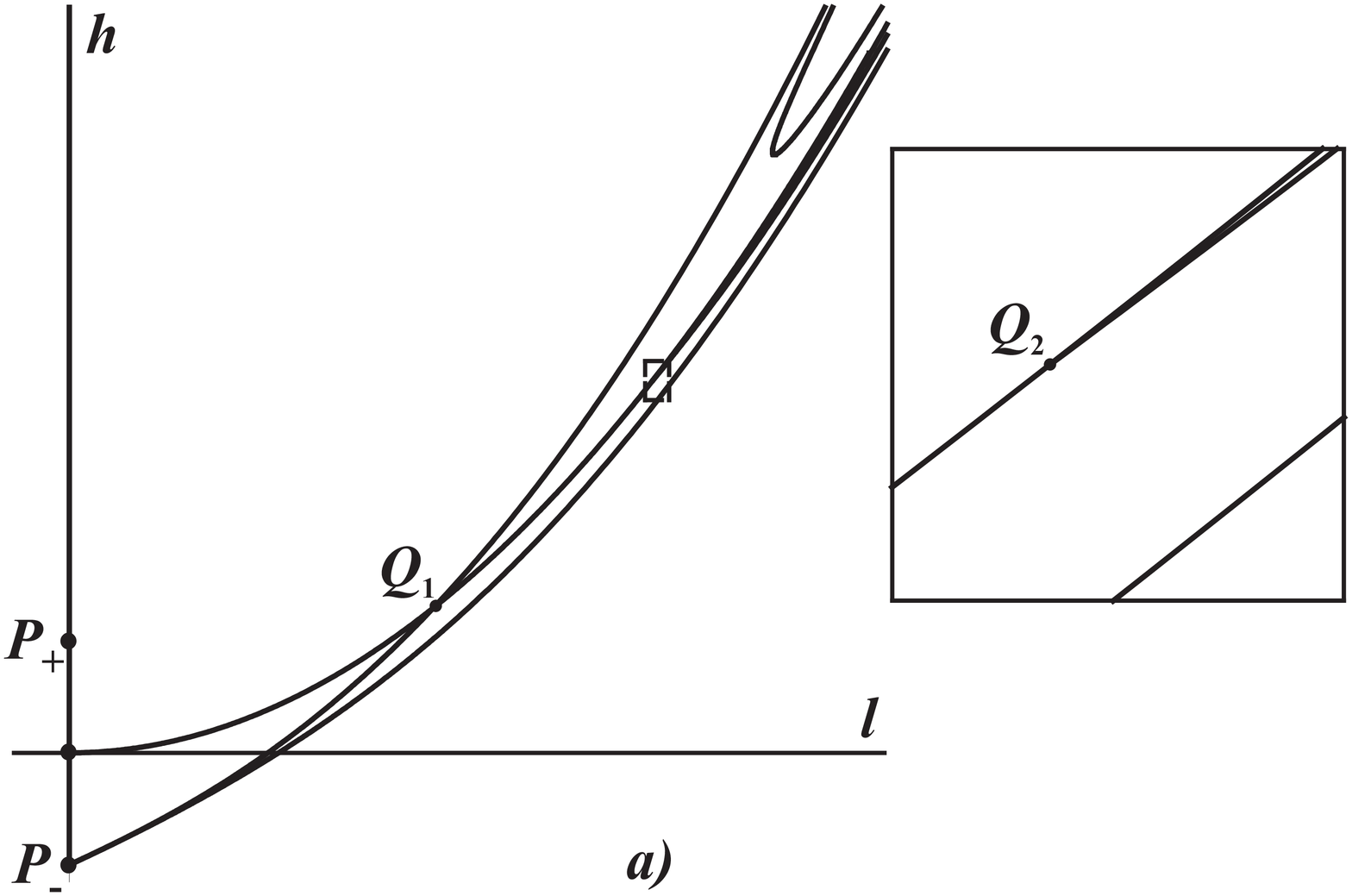}
\includegraphics[width=60mm,keepaspectratio]{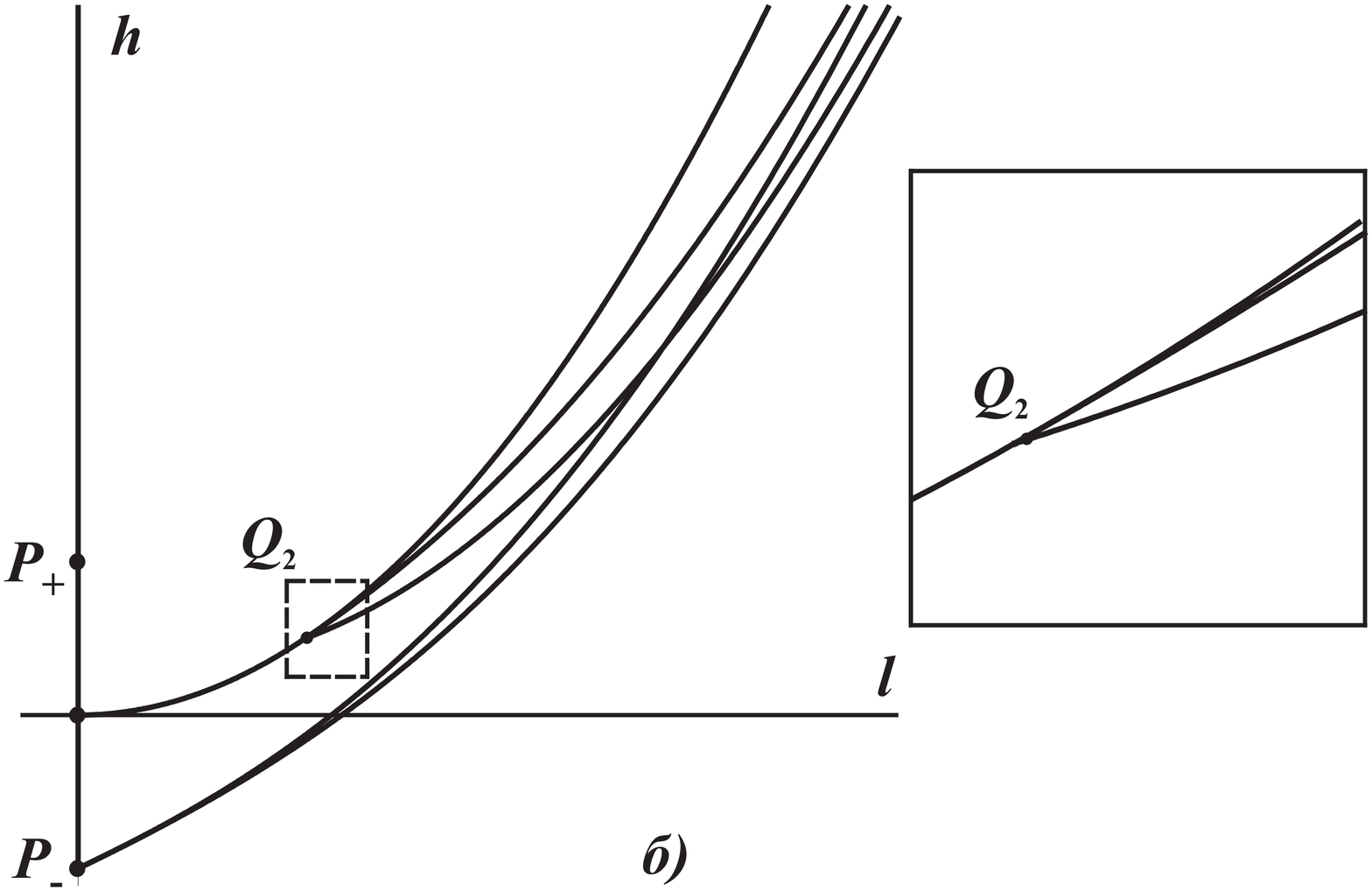}
\caption{Бифуркационные диаграммы при $n>2$.}
\end{figure}

Основные разделяющие значения $n$ можно найти аналитически.
Рассмотрим, например, точку $Q_1$ пересечения кривой $\delta_2$ с
первой ветвью кривой $\delta_4$ ($x \in [0,1)$). Она отмечена на
рис.~2, {\emph{а}}. Видно, что при переходе от выбранного случая
\emph{а}) к случаю \emph{б}) точка $Q_1$ исчезает. Соответствующее
разделяющее значение параметра $n$ обозначим через $n_*$. Для того
чтобы найти это значение, заметим, что параметр $x$ на кривой
$\delta_4$ в точке пересечения с $\delta_2$ есть корень многочлена
$P(x)=(n-1)^3x^3-2(n-1)(n+5)x^2+4(5n-2)x-8n$ на полуинтервале
$[0,1)$. Поскольку $P(0)=-8n$ и $P(1)=-(n+1)(n-3)$, пересечение
существует для всех $n<n_*=3$. Результант $P(x)$ и $P'(x)$ равен
$256(n-1)^4(n-3)(n^4-5n^3+18n^2+2n+11)$ и не обращается в нуль при
$n>3$. Следовательно, при $n>3$ многочлен $P(x)$ имеет
единственный вещественный корень, который всегда больше единицы.
Итак, точка $Q_1$ не может возникнуть снова для значений $n>3$.

На рис.~2 также отмечена всегда существующая при $n>2$ точка $Q_2$,
в которой встречаются кривые $\delta_2$, $\delta_5$ и вторая ветвь
кривой $\delta_4$. Она при некотором $n$ пересекает первую ветвь
кривой $\delta_4$. Очевидно, что при этом $Q_2=Q_1$. Координаты
$Q_2$ легко находятся из уравнений $\delta_4, \delta_5$ при $x=2$:
$$
l=\frac{4}{\sqrt{n-2}}, \qquad h=\frac{2}{n-2}.
$$
Предположим, что $Q_2 \in \delta_4$ при $x \ne 2$, исключим $x$ и
получим уравнение $n^4-3n^3-5n^2+20n-11=0$, которое имеет ровно
четыре вещественных корня. Разделяющий случай $Q_2=Q_1$ отвечает
наибольшему корню $n \approx 2.538$.

Итак, в дополнение к значению $n=2$, при котором система
интегрируема в целом, отмечены еще два значения $n$, когда
бифуркационные диаграммы терпят перестройки. В том числе, особым
значением является $n=3$. Возможно, что этому соответствуют
некоторые частные случаи интегрируемости.

\vskip2mm\Section[n]{Заключение}\label{sec6} В работе рассмотрено
однопараметрическое семейство механических систем с тремя
степенями свободы, обладающих $S^1$-симметрией с нетривиальным
множеством особых точек. Получены уравнения бифуркационных
диаграмм соответствующих отображений момента. Приведены некоторые
примеры перестроек диаграмм, показывающие, что полная
классификация всех диаграмм по физическому параметру $n$ может
оказаться нетривиальной задачей.

Из работ \cite{Yeh, ReySem, KhND} следует, что дальнейшие
обобщения имеют место для задачи о движении динамически
симметричного гиростата в двух постоянных полях. Семейство
диаграмм будет зависеть уже от двух физических параметров.

Если исключить из фазового пространство критическое интегральное
многообразие $L=0$, то оставшаяся система приводится к двум степеням
свободы. Конфигурационное пространство -- сфера с выколотой точкой
-- диффеоморфно $\R^2$. Приведенный потенциал выписать несложно, но
для нахождения интегральных многообразий необходимо вычислить
индексы его особых точек. Сами особые точки фактически найдены выше.
Топологический анализ выходит за рамки настоящей статьи и
планируется в дальнейшем для гиростата с интегралом Яхья.

Как уже отмечалось, собственно в случае Яхья ($n=2$) система
допускает еще один интеграл $K$, найденный в фундаментальной
работе О.И.~Богоявленского \cite{Bogo} и обобщающий интеграл
Ковалевской. В связи с этим, можно ставить задачу об исследовании
бифуркационной диаграммы возникающего отображения
\begin{equation}\label{eq6_1}
H \times L \times K: P^6 \to \R^3.
\end{equation}
Эта диаграмма должна получиться как вырождение общих диаграмм,
построенных для волчка Ковалевской--Реймана--Семенова-Тян-Шанского
\cite{ReySem} в работе \cite{Kh34}, с учетом возникающей связи
обобщения интеграла площадей с интегралами $H, L$. Отметим, что
для частного интеграла Богоявленского, указанного им на
инвариантном многообразии $M^4=\{K=0\}$ (в случае Яхья интеграл
Богоявленского совпадает с ограничением $L$ на многообразие
$M^4$), бифуркационная диаграмма отображения $H\times L$ и
топология системы с двумя степенями свободы на $M^4$ при условиях
Яхья изучены в работе \cite{Zotev}. Показано, что в этом случае
$M^4$ не является всюду гладким, и выявлены интегральные
поверхности с самопересечениями (погруженные многообразия).
Исследование отображения (\ref{eq6_1}) позволит получить описание
трехмерного слоения Лиувилля в случае Яхья, включающие найденные в
\cite{Zotev} нетривиальные бифуркации как бифуркации внутри
критических подсистем.

Работа выполнена при финансовой поддержке гранта РФФИ и
Администрации Волгоградской области № 10-01-97001.

\end{document}